\documentclass[sigconf]{acmart-me}

\usepackage{booktabs} % For formal tables
\usepackage{url}
\usepackage{hyperref}
\usepackage{color}
\usepackage{enumitem}
\setcopyright{rightsretained}

\acmDOI{}

\acmISBN{}

\acmConference[MediaEval'20]{Multimedia Evaluation Workshop}{14-15 December 2020}{Online} 
\acmYear{2020}
\copyrightyear{2020}

\acmPrice{}

\begin{document}

\title{Investigating Memorability of Dynamic Media}
%\titlenote{Produces the permission block, and
%  copyright information}
%\subtitle{Extended Abstract}
%\subtitlenote{The full version of the author's guide is available as
 % \texttt{acmart.pdf} document}

% \date{November 2020}

\author{Phuc H. Le-Khac, Ayush K. Rai, Graham Healy, Alan F. Smeaton, Noel E. O'Connor}
\affiliation{Dublin City University, Ireland}
\email{{khac.le2, ayush.rai3}@mail.dcu.ie}
\email{{graham.healy, alan.smeaton, noel.oconnor}@dcu.ie}

\renewcommand{\shortauthors}{P.H. Le-Khac et al.}
\renewcommand{\shorttitle}{Predicting Media Memorability}

\newcommand{\phuc}[1]{\textcolor{cyan}{\textbf{Phuc said:} #1}}
\newcommand{\alan}[1]{\textcolor{red}{\textbf{Alan said:} #1}}
\newcommand{\graham}[1]{\textcolor{blue}{\textbf{Graham said:} #1}}
\newcommand{\ayush}[1]{\textcolor{magenta}{\textbf{Ayush said:} #1}}

\begin{abstract}

The Predicting Media Memorability task in MediaEval'20 has some challenging aspects compared to  previous years.
In this paper we identify the high-dynamic content in videos and dataset of limited size as the core challenges for the task, we propose directions to overcome some of these challenges and we present our initial result in these directions.
\end{abstract}

%\keywords{ACM proceedings, \LaTeX, text tagging}

\maketitle

\section{Introduction}
\label{sec:intro}
Since the seminal paper on image memorability by Isola \emph{et. al.} \cite{isola2011UnderstandingIntrinsicMemorability}, there has been growing interest in building computational models to understand and predict the intrinsic memorability of media, as well as other subjective perceptions of media \cite{constantin2019ComputationalUnderstandingVisual}.
The Predicting Media Memorability task at MediaEval is designed to facilitate and promote research in this area by requiring task participants to develop  computational approaches to generate  measures of media memorability, which in turn also helps to better understand the subjective memorability of human cognition.
Having a prediction model of a media's memorability also allows for interesting applications to be developed, such as techniques to enhance memorability of images through the use of neural style transfer \cite{siarohin2019IncreasingImageMemorability}.

Continuing  from the success of the previous two years \cite{cohendet2018MediaEval2018Predicting, constantin2019PredictingMediaMemorability}, the MediaEval'20 Predicting Media Memorability challenge \cite{secodeherrera2020OverviewMediaEval2020} remains the same, where teams need to build prediction models for the   memorability scores from a dataset of videos with captions. 
%Memorability scores are measured as the ratio of the number of people that correctly recall having seen a video over the total number of participants that have been shown that video for a second time.
There are two types of memorability scores: short-term scores for videos that are shown for a second time within a short timespan of the initial viewing, and long-term scores for videos that are shown again two to three days later.
The videos making up this year's training dataset contain 590 videos with sound from the TRECVid 2019 Video-to-text dataset \cite{awad2020TRECVID2019Evaluation}, compared to 8,000 soundless videos from  VideoMem \cite{cohendet2019VideoMemConstructingAnalyzing} dataset.

\section{Related Work}
\label{sec:work}
In previous challenges, a  variety of methods that utilised different data modalities were explored \cite{tran2019PredictingMediaMemorability, leyva2019MultimodalDeepFeatures, azcona2019PredictingMediaMemorability}.
As previous works have shown \cite{constantin2019UsingAestheticsAction, reboud2019CombiningTextualVisual}, utilising high-level semantic features, either extracted using deep networks or provided by human annotators via text captions, are among the most effective methods to predict memorability.
However, the  modalities and  features that are most predictive of memorability remain unclear i.e. the best approach on last year's challenge \cite{azcona2019PredictingMediaMemorability} used an ensemble of models trained on a  variety of modalities.

\section{Approach}
\label{sec:approach}

\subsection{Motivation}

\begin{figure}[h]
    \includegraphics[width=\linewidth]{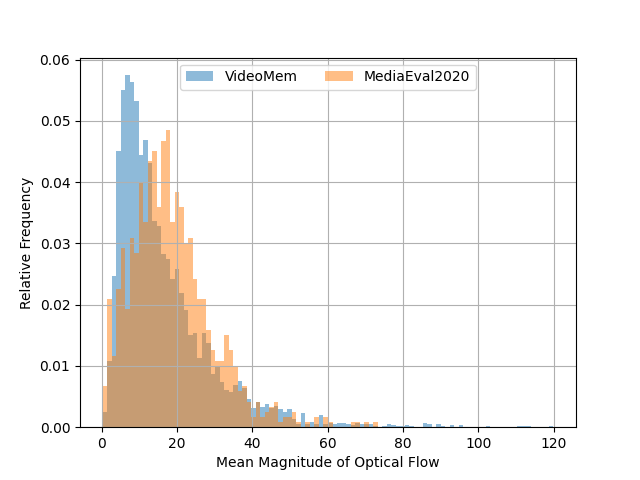}
    \caption{Histogram showing distributional differences in Mean Magnitude of Optical Flow (amount of motion) between the VideoMem dataset used in previous years and the MediaEval2020 dataset. 
    The Mean Magnitude of Optical Flow of a video was obtained by averaging the optical flow features over all pixels in a frame and across all frames.}
    \label{fig:flow}
\end{figure}

\paragraph{High dynamic video}
The short videos used in previous years' memorability challenges were extracted from raw professional footage that is used for example in creating high-quality film and commercials \cite{cohendet2019VideoMemConstructingAnalyzing}.
Consequently, the majority of videos therefore contain only one scene and are mostly static.
Conversely, the data for this year's 2020 challenge is extracted from the TRECVid 2019 \cite{awad2020TRECVID2019Evaluation} Video-to-Text dataset.
These videos were collected from social media posts and TV shows. They contain multiple scene changes in one video, and thus are more dynamic and with more complex movement and activity levels, as illustrated in Fig.\ref{fig:flow}.

While previous approaches to memorability computation use higher-semantic features, such as those extracted from deep neural networks, text captions provided by human-annotators and human-centric features such as emotions or aesthetics, most of these features were extracted using frame-based models.
In contrast, this year's dataset provided an unexplored challenge for methods that can capture semantic features from highly dynamic videos.

\paragraph{Small size dataset}
Another challenge for this year's data set is the limited number of annotated videos with memorability scores.
Compared to previous years, this is an order of magnitude smaller in size, with 590 videos in the training set and 500 videos in the test set. 
A development set with an additional 500 videos was released in the later stage of the benchmark.

\paragraph{Challenging benchmark}
The changes to this year's dataset made the task considerably more challenging, as the videos used were more dynamic thus making the extraction of high-level semantic features more difficult, and  further compounded by the limited size of the dataset.

Motivated by these insights, we focus our work on using video-level features to capture the dynamic in videos, and attempt to pre-train a memorability model on a larger dataset before fine-tuning it on this year MediaEval's data.

\subsection{Spatiotemporal baseline}
While the provided captions for each video can be used as high-level semantic-rich features for predicting media memorability. However, in our approach we focus on a method to extract high-level features from raw video input, without any additional annotation.

Motivated from the fact that this year's video is highly dynamic as discussed above, we chose to use features extracted from C3D \cite{tran2015LearningSpatiotemporalFeatures}, the only video-based method instead of frame-based model. We used the C3D pre-extracted features provided by the task organisers as the input for our memorability regression model.

\paragraph{C3D model learned spatio-temporal representation}
C3D \cite{tran2019PredictingMediaMemorability}, short for 3D Convolution, is among the earliest approaches to learning generic representation from videos.
Extending the 2D Convolution operations common in most image-processing deep learning models, C3D's homogeneous architecture composed of small $3 \times 3 \times 3$ convolution kernels, expanding in all height, width and time dimension of videos.
Trained on a generic action recognition dataset, each video passed through C3D returns a 4096-dimension feature vector that extracts high-level semantic features from a video segment.

\paragraph{Memorability Regression Model}
With the extracted features via C3D, we used a multi-layer perceptron (MLP) with two hidden layers of 512 units followed by a Rectified Linear Unit (ReLU) non-linearity layer.
To mitigate overfitting, we used Dropout \cite{srivastava2014DropoutSimpleWay} with a probability of 0.1.
Batch Normalisation \cite{ioffe2015BatchNormalizationAccelerating} was used to normalise the features in each batch after each layer to help optimisation.
The final model is a sequential application 3 building blocks composed from BatchNorm, Dropout, fully-connected and ReLU activation layers.
For the final output layer, the ReLU activation is replaced with a Sigmoid to obtain the final memorability score in the range from 0 to 1.

There was a total of more than 2 million parameters in the model, most belonging to the first fully-connected layer due to the large C3D's feature size of 4096.
To minimise the effect of outliers in ranking score, we use L1 loss instead of the standard L2 Mean-Squared Error (MSE) and saw a slight improvement in convergence speed.
The model is trained for 100 epochs on the training set and the final model was picked based on the performance on the validation set.

% We could train on the VideoMem dataset of last year challenge, however due to the drastic change in the nature of videos, the pre-training may not be very effective on this year dataset.
\subsection{Large scale Memorability Pre-training}
To overcome the limited size of this year's dataset, we used the recently released Memento10K \cite{newman2020MultimodalMemorabilityModeling} for pre-training our model before fine tuning on the MediaEval challenge's dataset. Containing 10,000 videos, the Memento10K dataset is roughly the same size as the VideoMem \cite{cohendet2019VideoMemConstructingAnalyzing} dataset.
Moreover, videos in the Memento10K contains more action and are more similar to the new data of this year's challenge.  
Therefore we decided to focus our large-scale pre-training approach on the Memento10K dataset and replicate the accompanying SemNet model \cite{newman2020MultimodalMemorabilityModeling} instead of pre-training on VideoMem \cite{cohendet2019VideoMemConstructingAnalyzing} from previous years.
The SemNet model contains three separate sub-networks to process three different input streams: \textit{image}, \textit{optical flow} and \textit{video} stream.

After pre-training SemNet on Memento10k data, we only retain the video stream sub-network to fine tune on the MediaEval dataset and discarded all other components.

\section{Results and Discussion}
The results of our runs together with this year's mean and variance are reported in Table~\ref{tab:result}. 
Due to technical issues and time constraints, we only submitted the baseline regression model based on C3D features and did not have the result for fine tuning SemNet on the Memento10K \cite{newman2020MultimodalMemorabilityModeling} dataset for this year's challenge.

\begin{table}[h]
  \caption{Results of our approach compared with the challenge's statistics}
  \label{tab:result}
  %\scalebox{\0.9}{%
  \resizebox{\columnwidth}{!}{
    \begin{tabular}{@{}ccccccc@{}}
    \toprule
        & \multicolumn{3}{c}{Short-term} & \multicolumn{3}{c}{Long-term} \\
    \midrule
    Run & \textbf{Spearman} & Pearson & MSE & \textbf{Spearman} & Pearson & MSE \\ 
    \midrule
    Mean & 0.058 & 0.066 & 0.013 & 0.036 & 0.043 & 0.051 \\
    Variance & 0.002 & 0.002 & 0.000 & 0.002 & 0.001 & 0.000 \\ 
    \midrule
    \textbf{C3D} & 0.034 & 0.078 & 0.1 & -0.01 & 0.022 & 0.09 \\ 
    \bottomrule
\end{tabular}
}
\end{table}

\noindent 
Even with the simplest method possible, the result from our regression baseline with  pre-extracted C3D features is not very far  from the mean.
This support our hypothesis that spatio-temporal representation is important for this year's dataset and the performance can be increased with other complementary high-level features.
Future work can explore this hypothesis further by fine tuning the entire C3D model instead of just using the pre-extracted features, or by using different models that also capture spatio-temporal features.

On the other hand, compared to the state-of-the-art result of $0.528$ \cite{azcona2019PredictingMediaMemorability} in last year's challenge, the mean of the Spearman rank correlation for all runs this year ($0.058)$ is an order of magnitude lower. This highlights the importance of large scale pre-training and transfer learning techniques, and methods that can extract high-level features from dynamic videos effectively.

% The dramatic change of performance from last year indicates a step forward in the progress of studying the memorability of media.
% It provides many challenges, video contains more dynamic, and has to overcome the limited dataset issues.

%Even though our initial experiments in this year challenge has not been satisfactory, 
%We believe that the problems that we identified and proposed solutions are pointing in the right direction, and with more thorough experiments, it can achieve better results on this challenging challenge.

We believe the direction of our solution, given the challenges in this year's challenge, not only shows promise but also indicates the importance of spatio-temporal models to capture high-level semantics of videos for memorability prediction.

\begin{acks}
This work was co-funded by Science Foundation Ireland through the SFI Centre for Research Training in Machine Learning (18/CRT/6183) and the Insight Centre for Data Analytics (SFI/12/RC/2289\_P2), co-funded by the European Regional Development Fund. A.K. Rai also acknowledges support from FotoNation Ltd.
\end{acks}

\bibliographystyle{ACM-Reference-Format}
\def\bibfont{\small} % comment this line for a smaller fontsize
\bibliography{main.bib} 

\end{document}